\documentclass{llncs}

\usepackage{txfonts}

\usepackage{mathtools} 
\usepackage{stmaryrd} 
\usepackage[algoruled,vlined,english]{algorithm2e}

\usepackage{listings}
\usepackage{galois}

\usepackage{paralist}

\usepackage{todonotes}

\usepackage[pdftex,%
 	colorlinks,%
	hidelinks,
 	pdfauthor={Pierre Ganty, Samir Genaim},%
 	hypertexnames=false,%
	pdftitle={Proving Termination Starting from the End}]{hyperref}

\newcommand{\tuple}[1]{\langle #1 \rangle}
\newcommand{\goodtr}[1]{\ensuremath{{#1}_{G}}}
\newcommand{\badtr}[1]{\ensuremath{{#1}_{B}}}
\newcommand{\pre}{\ensuremath{\mathit{pre}}}
\newcommand{\post}{\ensuremath{\mathit{post}}}

\newcommand{\lfp}{\ensuremath{\mathit{lfp}}}
\newcommand{\gfp}{\ensuremath{\mathit{gfp}}}

\newcommand{\anyinput}{\ensuremath{\mathit{true}}\xspace}
\newcommand{\linear}{\ensuremath{\star}\xspace}
\newcommand{\optimal}{\ensuremath{\bullet}\xspace}
\newcommand{\acabar}{\texttt{Acabar}\xspace}

\title{Proving Termination Starting from the End}
\author{Pierre Ganty\inst{1} \and Samir Genaim\inst{2}}
\institute{IMDEA Software Institute, Madrid, Spain \and Universidad Complutense de Madrid, Spain}

\begin{document}
\pagestyle{plain}

\maketitle

%
%

\begin{abstract}
	We present a novel technique for proving program termination which introduces
	a new dimension of modularity.
	Existing techniques use the program to incrementally construct a termination
	proof. While the proof keeps changing, the program remains the same.
	Our technique goes a step further. We show how to use the current partial
	proof to partition the transition relation into those behaviors known to be
	terminating from the current proof, and those whose status (terminating or
	not) is not known yet. This partition enables a new and unexplored dimension
	of incremental reasoning on the program side.
	In addition, we show that our approach naturally applies to conditional
	termination which searches for a precondition ensuring termination.
	We further report on a prototype implementation that advances the
	state-of-the-art on the grounds of termination and conditional termination.  
%
\end{abstract}

%
%

\section{Introduction}

The question of whether or not a given program has an infinite execution is a
fundamental theoretical question in computer science but also a highly
interesting question for software practitioners.  The first major result is
that of  Alan Turing, showing  that the \emph{termination problem} is 
undecidable.
Mathematically, the termination problem for a given program
\(\mathit{Prog}\) is equivalent to deciding whether the transition
relation \(R\) induced by \(\mathit{Prog}\) is well-founded.


The starting point of our paper, is a result showing
that the well-foundedness problem of a given relation \(R\) is equivalent to
the problem of asking whether the transitive closure of \(R\), noted \(R^+\), 
is disjunctively well-founded \cite{PR04}. That is whether \(R^+\) is included in some \(W\)
(in which case \(W\) is called a \emph{transition invariant}) such that \(W= W_1 \cup \cdots \cup W_n \), \(n\in\mathbb{N}\) and each \(W_i\) is
well-founded (in which case \(W\) is said to be \emph{disjunctively
well-founded}). This result has important practical consequences because it
triggered the emergence of effective techniques, based on transition invariants, to solve the termination
problem for real-world
programs~\cite{CookPR06,AlbertAGPZ07,SpotoMP10,KroeningSTW10}.

By replacing the well-foundedness problem of \(R\) with the equivalent disjunctive
well-foundedness problem of \(R^+\), one allows for the incremental construction
of \(W\): when the inclusion of \(R^+\) into \(W\) fails then use the information
from the failure to update \(W\) with a further well-founded relation~\cite{CookPR05}.
Although the proof is incremental for \(W\), it is important to note that a similar
result does not hold for \(R\). That is, it is in general not true that given
\(R=R_1\cup R_2\), if \(R_1^+\subseteq W\) and \(R_2^+ \subseteq W\) then \(R^+
\subseteq W\). 

We introduce a new technique that, besides being incremental for \(W\), further
partitions the transition relation \(R\) separating those behaviors known to be
terminating from the current \(W\), from those whose status (terminating or
not) is not known yet. 
Formally, given \(R\) and a candidate \(W\), we shall see how
to compute a partition \(\{\goodtr{R}, \badtr{R}\}\) of \(R\)
such that
\begin{inparaenum}[\upshape(\itshape a\upshape)]
	\item \(\goodtr{R}^+ \subseteq W\); and
	\item every infinite sequence \(s_1 \mathbin{R} s_2 \mathbin{R} \cdots s_i \mathbin{R} s_{i+1}\cdots\)\linebreak (or \emph{trace}) has a
  suffix that exclusively consists of transitions from
  \(\badtr{R}\), namely we have \linebreak \(s_z \mathbin{\badtr{R}} s_{z+1} \mathbin{\badtr{R}} \cdots \) for some \(z\geq 1\). 
\end{inparaenum}

It follows that well-foundedness of \(\badtr{R}\) implies that of \(R\).
Consequently, we can focus our effort
exclusively on proving well-foundedness of \(\badtr{R}\).
In the affirmative, then so is
\(R\) and hence termination is proven. In the negative, then we have found an
infinite trace in \(\badtr{R}\), hence in \(R\). 
We observed that working with \(\badtr{R}\) typically provides further
hints on which well-founded relations to add to \(W\).
The partition of \(R\) into \(\{\goodtr{R},\badtr{R}\}\) enables a new and unexplored
dimension of modularity for termination proofs. 

Let us mention that the partitioning of \(R\) is the result of adopting a fixpoint
centric view on the disjunctive well-foundedness problem and leverage
equivalent formulation of the inclusion check.  
More precisely, we
introduce the dual of the check \( R^+ \subseteq W\) by defining the adjoint to
the function \(\lambda X\ldotp X\comp R\) used to define \(R^+\). Without
defining it now, we write the dual check as follows: \( R \subseteq W^-\).
%
We shall see that while the failure of \( R^+ \subseteq W\) provides
information to update \(W\); the failure of \(R \subseteq W^{-}\) provides
information on \emph{all pairs} in \(R\) responsible for the failure of \(W\)
as a transition invariant. This is exactly that information, of 
semantical rather than syntactical nature, that we use to
partition \(R\).

We show that the partitioning of \(R\) can be used not only for
termination, but it also serves for conditional termination. The goal
here is to compute a precondition, that is a set
\(\mathcal{P}\) of states, such that no infinite trace starts from a state of
\(\mathcal{P}\).  We show how to compute a (non-trivial) precondition from the
relation \(\badtr{R}\).

Our contributions are summarized as follows:
\begin{inparaenum}[\upshape(\itshape i\upshape)]
	\item we present \texttt{Acabar}, a new algorithm which allows for enhanced modular
		reasoning about infinite behaviors of programs;
	\item we show that, besides termination, \texttt{Acabar} can be used in the
		context of conditional termination; and
	\item finally, we report on a prototype implementation of our techniques and compare
it with the state-of-the-art on two grounds: the termination problem, and the
problem of inferring a precondition that guarantees termination.
\end{inparaenum}

\section{Example}\label{sec:motivating}

In this section, we informally overview our proposed techniques on an
example taken from the literature~\cite{CookGLRS08}. 
Consider the following loop:
\begin{lstlisting}
	while ( x>0 ) { x:=x+y; y:=y+z; }
\end{lstlisting}
represented by the transition relation \( R=\{ x > 0, x'=x+y, y'=y+z,
z'=z\} \), where the primed variables represent the values of the
program variables after executing the loop body.
Note that, depending on the input values, the program may not
terminate (e.g. for \(x=1\), \(y=1\) and \(z=1\) ). Below we apply \acabar  to prove termination. 
As we will see, this attempt ends with a failure which provide
information on which subset of the transition relation to blame.
Then, we will explain how to compute a termination
precondition from this subset.
%

In order to prove termination of this loop, we
seek a disjunctive well-founded relation \(W\) such that \(R^+ \subseteq W\).
To find such a \(W\), \acabar is supported by incrementally (and
automatically) inferring (potential) linear ranking functions for \(R\) or
\(R^+\)~\cite{CookGLRS08,CookPR05}.
When running on \(R\),  \acabar first adds the
candidate well-founded relation \( W_1=\{x'<x, x > 0\} \) to \(W\) which is initially
empty. Relation \(W_1\) stems from the observation that, in \(R\), \(x\) is
bounded from below (as shown by the guard) but not necessarily decreasing. Hence, using \(W=W_1\),
\acabar partitions \(R\) into \(\{\goodtr{R}^{(1)},\badtr{R}^{(1)}\}\) where:
\[
\begin{array}{rl}
\goodtr{R}^{(1)}=&\{x>0,x'=x+y,y'=y+z,z'=z, y<0, z\le 0\} \\[3pt]
 \badtr{R}^{(1)}=& \{x>0, x'=x+y, y'=y+z, z'=z,  y<0,  z> 0\}\lor\\
                & \{x>0, x'=x+y, y'=y+z, z'=z, y\geq0\}\enspace .
\end{array}
\]
The partition comes with the further guarantee that \emph{every infinite trace in \(R\) must have a
  suffix that exclusively consists of transitions from
  \(\badtr{R}^{(1)}\)}, which means that if \(\badtr{R}^{(1)}\) is
well-founded then so is \(R\).
In addition, one can easily see that \((\goodtr{R}^{(1)})^+ \subseteq
W\).

Next, \acabar calls itself recursively on \(\badtr{R}^{(1)}\) to show its
well-foundedness.
As before, it first adds \(W_2=\{y'<y,y\geq 0\}\) to \(W\). Similarly to
the construction of \(W_1\), \linebreak \(W_2\) stems from the observation that, in some
parts of \(\badtr{R}^{(1)}\), \(y\) is bounded from below but not
necessarily decreasing.
%
Then, using \(W=W_1\lor W_2\), \acabar partitions \(\badtr{R}^{(1)}\) into:
\[
\begin{array}{rl}
\goodtr{R}^{(2)}=&\{x>0,x'=x+y,y'=y+z,z'=z, z <0\}\\[3pt]
 \badtr{R}^{(2)}=&\{x>0,x'=x+y,y'=y+z,z'=z, y\geq 0,z\ge0\}\enspace .
\end{array}
\]
Again the partition \(\{\goodtr{R}^{(2)}, \badtr{R}^{(2)}\}\) of \(\badtr{R}^{(1)}\) comes with a similar guarantee. This time it holds that that every infinite
trace in \(R\) must have a suffix that exclusively consists of
transitions from \(\badtr{R}^{(2)}\).
Recursively applying \acabar on \(\badtr{R}^{(2)}\) does not yield any further
partitioning, that is \(\badtr{R}^{(3)}=\badtr{R}^{(2)}\). 
The reason being that no potential ranking function is automatically inferred.
Thus, \acabar fails to prove well-foundedness of \(R\), which is indeed not well-founded.
However, due to the above guarantee, we can use \(\badtr{R}^{(2)}\) to infer a
sufficient precondition for the termination of \(R\). We explain this next.

Inferring a sufficient precondition is done in two steps:
\begin{inparaenum}[\upshape(\itshape i\upshape)]
\item we infer (an overapproximation of) the set of all states
	\(\mathcal{Z}\) visited by some infinite sequence of steps in \(\badtr{R}^{(2)}\); and
\item we infer (an overapproximation of) the set of all states
	\(\mathcal{V}\) each of which can reach \(\mathcal{Z}\) through some steps in
	\(R\).
\end{inparaenum}
Turning to the example, we infer \(\mathcal{Z}=\{x>0,y \geq 0,z \geq0\}\) and
the following overapproximation \(\mathcal{V}'\) of \(\mathcal{V}\):
\[
\mathcal{V}'=\{x\geq 1,z=0,y\geq 0\} \lor \{x\geq 1,z\geq 1,x+y\geq
1,x+2y+z\geq 1,x+3y+3z\geq 1\}\enspace .
\]
It can be seen that every infinite trace visits only states in \(\mathcal{V}'\), hence the complement of \(\mathcal{V}'\) is a precondition for termination.

Let us conclude this section by commenting on an example for which \acabar
proves termination. Assume that we append \lstinline!z:=z-1! to the loop body
above and call \(R'\) the induced transition relation. Following our
previous explanations, running \acabar on \(R'\) updates \(W\) from
\(\emptyset\) to \(W_1\), and then to \(W_1\lor W_2\). Then, and contrary to the
previous explanations, \acabar will further update \(W\) to \(W_1\lor W_2\lor
W_3\) where \(W_3\) is the well-founded relation \(\{z'<z, z\ge0\}\). 
From there, \acabar returns with value \(\badtr{R}^{(3)}=\emptyset\), %
hence we have that \(R'\) is well-founded.

%

\section{Preliminaries}

A \emph{transition system} is a pair $(\mathcal{Q},R)$ where
$\mathcal{Q}$ is the set of \emph{states} and $R\subseteq \mathcal{Q}\times
\mathcal{Q}$ is the \emph{transition relation}.  An \emph{initialized}
transition system includes a further component $\mathcal{I}\subseteq
\mathcal{Q}$, the set of \emph{initial states}.
For simplicity, we defer the treatment of initial states to Sec.~\ref{sec:conclusion}.

An \(R\)-\emph{trace} is a sequence \(s_1,s_2,\ldots, s_n\) of states such that
for every \(i\), \(1\leq i < n\) we have \( (s_i,s_{i+1})\in R\).  When \(R\)
is clear from the context we simply say trace.  An \emph{infinite \(R\)-trace}
is a sequence \(s_1,s_2,\ldots\) of states such that for every \(i\geq 1\) we
have \( (s_i,s_{i+1})\in R\). Given \(R'\subseteq R\) and an infinite
\(R\)-trace \(\pi\) we say that \(\pi\) has \emph{infinitely many steps} in
\(R'\) if \( (s_i,s_{i+1})\in R'\) for infinitely many \(i\geq 1\).


Given a relation $R' \subseteq R$ and a set $\mathcal{Q}' \subseteq
\mathcal{Q}$, define $\post[R'](\mathcal{Q}')\stackrel{\mathrm{def}}{=}\{ s'
\in \mathcal{Q} \mid \exists s \in \mathcal{Q}' \colon (s,s') \in R' \}$.  We
say that this operator computes the \emph{\(R'\)-successors of \(Q'\)}.
Dually, define
$\pre[R'](\mathcal{Q}')\stackrel{\mathrm{def}}{=}\post[R'^{-1}](\mathcal{Q}')=\{
s \in \mathcal{Q} \mid \exists s' \in \mathcal{Q}' \colon (s,s') \in R' \}$.
We say that this operator computes the \emph{\(R'\)-predecessors of \(Q'\)}.%
\footnote{We define \(R^{-1}, R^{*}\) and \(R^{+} \) to be \( R^{-1}=\{(s',s) \mid (s,s')\in R\}\), \(R^{*}=\bigcup_{i\geq 0} R^i\) and \(R^{+}=R\comp R^{*}\) where
\(R^0\) is the identity, \(R^{i+1}= R^i\comp R\) and
\(R_1\comp R_2 = \{ (s,s'') \mid \exists s'\colon (s,s')\in R_1\land (s',s'')\in R_2\}\).}

A relation \(W \subseteq \mathcal{Q}\times \mathcal{Q}\) is called
\emph{disjunctively well-founded} if{}f \(W\) coincides with the union of
finitely many relations (viz. \(W= W_1 \cup \ldots \cup W_n\)) each of which is
well-founded (viz. there is no infinite sequence \(s_1,s_2,\ldots\) such that
\( (s_{i},s_{i+1})\in W_{\ell}\) for all \(i\geq 1\)). 

In this paper, we adhere to the following conventions: calligraphic letters
\(\mathcal{X},\mathcal{Y},\ldots\) refer to subsets of \(\mathcal{Q}\)
and capital letters \(X,Y,\ldots\) refer to relations over \(\mathcal{Q}\), that is
subsets of \(\mathcal{Q}\times\mathcal{Q}\). Further, throughout the paper the
letter
\(W\) is used to denote a relation over \(\mathcal{Q}\) that is
disjunctively well-founded.

A \emph{linear expression} is of the form $a_0+a_1x_1+\cdots+a_nx_n$ where
$a_i\in\mathbb{Z}$ and $\bar{x}=\tuple{x_1,\ldots,x_n}$ are \emph{variables}
ranging over $\mathbb{Z}$. An \emph{atomic linear constraint} $c$ is of the
form $e_1~\mathit{op}~e_2$ where $e_i$ is a linear expression and
$op\in\{=,\geq,\leq,>,<\}$.
A \emph{formula} $\psi$ is a Boolean combination of atomic linear
constraints. Note that $\neg\psi$ is also a formula.
For the sake of simplicity, a conjunction $c_1\land \cdots\land c_n$
of atomic linear constraints is sometimes written as the set
$\{c_1,\ldots,c_n\}$.
A \emph{solution} of a formula $\psi$ is a mapping from its variables into
the integers such that the formula evaluates to true. 
%
Sets and relations over, respectively, \(\mathbb{Z}^n\) and
\(\mathbb{Z}^n\times\mathbb{Z}^n\) are sometimes specified using
formulas, with the customary convention, for relations, of variables
and primed variables. For instance, the formula $\{ x \geq 0, x'=x-y, y'=y \}$ defines
the relation \(R\subseteq \mathbb{Z}^2\times\mathbb{Z}^2\) such
that \(R=\{ \tuple{ (x,y),(x',y')} \mid x \geq 0 \land x'=x-y \land y'=y\}\).

Finally, we briefly recall classical results of lattice theory and refer to the
classical book of Davey and Priestley \cite{DaveyPriestley1989} for further
information.  Let $f$ be a function over a partially ordered set $(L,\sqsubseteq)$. A \emph{fixpoint} of $f$
is an element $l\in L$ such that $f(l)=l$.  We denote by
$\lfp\ f$ and $\gfp\ f$, respectively, the
\emph{least} and the \emph{greatest fixpoint}, when they exist, of $f$.  The
well-known Knaster-Tarski's theorem states that each order-preserving function $f\in
L\rightarrow L$ over a complete lattice
$\tuple{L,\sqsubseteq,\bigsqcup,\bigsqcap,\top,\bot}$ admits a least (greatest) fixpoint and
the following characterization holds: 
\begin{align}
  \lfp\ f&=\textstyle{\bigsqcap}\{x\in L \mid f(x)\sqsubseteq x\} & \gfp\ f &= \textstyle{\bigsqcup}\{x \in L \mid x \sqsubseteq f(x)\}\enspace .
\label{eq:charfp}
\end{align}

%

\section{Modular Reasoning For Termination}
\label{sec:modular}

A termination proof based on transition invariants consists in establishing the
existence of a disjunctively well-founded transition invariant. That is, the
goal is to prove the inclusion of \(R^{+}\), into some
\(W\).\footnote{Recall that \(W\) is always assumed to be disjunctively well-founded.} For short, we write  \(R^+ \subseteq W\).  Proving termination is thus
reduced to finding some \(W\) and prove that the inclusion hold. 

In the above inclusion check, \(R^+\) coincides with the least fixpoint of the
function \( \lambda Y \ldotp R \cup g(Y) \) where \(g\stackrel{\mathrm{def}}{=}\lambda Y \ldotp Y \comp R\).
It is known \cite{Cousot00-SARA} that if we can find an \emph{adjoint function}
\(\tilde{g}\) to \(g\)  such that \(g(X) \subseteq
Y\) if{}f \( X \subseteq \tilde{g}(Y)\) for all \(X,Y\) then there exists an
equivalent inclusion check to \(R^+ \subseteq W\).  This equivalent check,
denoted \(R\subseteq W^-\) in the introduction, is such that \(W^{-}\) is
defined as a greatest fixpoint of the function \(\lambda Y\ldotp W\cap
\tilde{g}(Y)\).
Next, we define \(\tilde{g}\stackrel{\mathrm{def}}{=}\lambda Y\ldotp \neg( \neg
Y\comp R^{-1})\).

\begin{lemma}
	Let \(X, Y\) be subsets of \(\mathcal{Q}\times \mathcal{Q}\) we have:
	\( X\comp R \subseteq Y \Leftrightarrow X \subseteq \neg( \neg Y\comp R^{-1})\).
	\label{lem:galois_comp}%
\end{lemma}
\begin{proof}
	First we need an easily proved logical equivalence:\\
	\hspace*{\stretch{1}}\((\varphi_1\land \varphi_2)\Rightarrow \varphi_3\) if{}f \(( \neg \varphi_3\land\varphi_2) \Rightarrow \neg \varphi_1\)\enspace .\hfill\vspace{0pt}\\
	Then we have:
	\begin{align*}
	             & X\comp R \subseteq Y	\\
	\text{if{}f } & \forall s,s',s_1 \colon \bigl( (s,s_1)\in X \land (s_1,s')\in R\bigr) \Rightarrow (s,s')\in Y\\
	\text{if{}f } & \forall s,s',s_1 \colon \bigl(  (s,s')\notin Y \land (s_1,s')\in R  \bigr) \Rightarrow (s,s_1)\notin X &\text{by above equivalence} \\
	\text{if{}f } & \forall s,s',s_1 \colon \bigl( (s,s')\notin Y \land (s',s_1)\in R^{-1}  \bigr) \Rightarrow (s,s_1)\notin X &\text{def.\ of }R^{-1}\\
	\text{if{}f } & \forall s,s',s_1 \colon \bigl( (s,s')\in \neg Y \land (s',s_1)\in R^{-1}  \bigr) \Rightarrow (s,s_1)\in \neg X \\
	\text{if{}f } & (\neg Y\comp R^{-1}) \subseteq \neg X\\
	\text{if{}f } & X\subseteq \neg(\neg Y\comp R^{-1}) & \text{\qed}
	\end{align*}
\end{proof}

Intuitively, \(g\) corresponds to \emph{forward reasoning} for proving
termination while \(\tilde{g}\) corresponds to \emph{backward reasoning}
because of the composition with \(R^{-1}\).  The least fixpoint \(\lfp\ \lambda
Y\ldotp R\cup g(Y)\) is the least relation \(Z\) containing \(R\) and closed by
composition with \(R\), viz. \(R\subseteq Z\) and \(Z\comp R\subseteq Z\).  On
the other hand, the greatest fixpoint \(\gfp\ \lambda Y\ldotp W\cap
\tilde{g}(Y)\) is best understood as the result of removing from \(W\) all
those pairs \( (s,s') \) of states such that \( (s,s')\comp R^+ \nsubseteq W\).
This process returns the largest subset \(Z'\) of \(W\) which is closed by composition
with \(R\), viz. \(Z'\subseteq W\) and \(Z'\comp R\subseteq Z'\). 
Using the results of Cousot~\cite{Cousot00-SARA} we find next that termination
can be shown by proving either inclusion of Lem.~\ref{lem:forback}.


\begin{lemma}[from \cite{Cousot00-SARA}] 
  \( \lfp\ \lambda Y\ldotp R\cup g(Y) \subseteq W \Leftrightarrow R \subseteq \gfp\ \lambda Y\ldotp W\cap \tilde{g}(Y) \).
\label{lem:forback}
\end{lemma}
\begin{proof}
	\begin{align*}
	  \lfp\ \lambda Y\ldotp R\cup g(Y) \subseteq W \text{ if{}f }& \exists A\colon R \subseteq A\land g(A) \subseteq A \land A \subseteq W &\text{by \eqref{eq:charfp}}\\
	  \text{if{}f }& \exists A\colon R\subseteq A\land A \subseteq \tilde{g}(A) \land A \subseteq W &\text{Lem.~\ref{lem:galois_comp}}\\
		\text{if{}f }& R\subseteq \gfp\ \lambda Y\ldotp W\cap \tilde{g}(Y) &\text{by \eqref{eq:charfp}} &	\text{\qed }%
 	\end{align*}%
\end{proof}

As we shall see, the inclusion check based on the greatest fixpoint has
interesting consequences when trying to prove termination.

An important feature when proving termination using transition invariants is to
define actions to take when the inclusion check \( \lfp\ \lambda Y\ldotp R\cup
g(Y) \subseteq W \) fails. In this case, some information is
extracted from the failure (e.g., a counter example), and is used to enrich $W$
with more well-founded relations~\cite{CookPR05}.

We shall see that, for the backward approach, failure of \(R \subseteq \gfp\
\lambda Y\ldotp W\cap \tilde{g}(Y) \) induces a partition of the transition
relation \( R \) into \( \{\goodtr{R}, \badtr{R}\} \) such that 
\begin{inparaenum}[\upshape(\itshape a\upshape)]
\item \( (\goodtr{R})^+ \subseteq W\); together with the following termination guarantee
\item every infinite \(R\)-trace contains a suffix that is an infinite
  \(\badtr{R}\)-trace (Lem.~\ref{lem:periodic}).  
\end{inparaenum}
An important consequence of this is that we can focus our effort
exclusively on proving termination of \(\badtr{R}\).
It is important to note that the guarantee that no infinite \(R\)-trace
contains infinitely many steps from \(\goodtr{R}\) is not true for any
partition \(\{\goodtr{R},\badtr{R}\}\) of \(R\) but it is true for our
partition which we define next.

\begin{definition}
  Let \( G = \gfp\ \lambda Y\ldotp W\cap \tilde{g}(Y)\), we define \(\{\goodtr{R},\badtr{R}\}\) to be the partition of \(R\) given by \( \goodtr{R}=R \cap G \) and \(
  \badtr{R}=R\setminus \goodtr{R} \).
  \label{def:partition}
\end{definition}

\begin{example}
  Let \(R=\{ x \ge 1, x'=x+y, y'=y-1\}\) and
  assume \(W=\{ x'<x, x\geq 1\}\) which is well-founded, hence disjunctively
  well-founded as well.
  Evaluating the greatest fixpoint (we omit calculations) yields
  \begin{align*}
  \goodtr{R}&=\{x\geq1,x'=x+y,y'=y-1,y<0\}\\
  \badtr{R}&=\{x\geq1,x'=x+y,y'=y-1,y\geq0\}
  \end{align*}
  which is clearly a partition of \(R\). The relation \(\goodtr{R}\) consists of those pairs
  of states where \(y\) is negative, hence \(x\) is decreasing as captured by
  \(W\).  On the other hand, \(\badtr{R}\) consists of those pairs where \(y\)
  is positive or null.  It follows that, when taking a step from \(\badtr{R}\),
  \(x\) does not decrease.  This is precisely for those pairs that \(W\) fails
  to show termination.\hfill\(\blacksquare\)
  \label{ex:firstrb}
\end{example}

Next, we state and prove the termination guarantees of the partition
\(\{\goodtr{R},\badtr{R}\}\).

\begin{lemma} 
  Given \(\goodtr{R}\) as in Def.~\ref{def:partition} we have \( \lfp\ \lambda Y\ldotp \goodtr{R} \cup Y \comp R \subseteq W\).
  \label{lem:goodtrwff}
\end{lemma}
\begin{proof}
\begin{align*}
	& G \subseteq \tilde{g}(G) \land G \subseteq W&\text{def.\ of \(G\) and \eqref{eq:charfp}}\\
	\text{only if }& g(G) \subseteq G \land G \subseteq W &\text{Lem.~\ref{lem:galois_comp}}\\
	\text{only if }& R\cap G \subseteq G \land g(G) \subseteq G \land G \subseteq W\\
	\text{only if }& \goodtr{R} \subseteq G \land g(G) \subseteq G \land G \subseteq W &\text{def.\ of \(\goodtr{R}\)}\\
	\text{only if }& \lfp\ \lambda Y\ldotp \goodtr{R} \cup g(Y) \subseteq W&\text{by \eqref{eq:charfp}}&\text{\qed}
\end{align*}
\end{proof}

An equivalent formulation of the previous result is \(\goodtr{R}\comp R^* \subseteq W\), which
in turn implies, since \(\goodtr{R}\subseteq R\), that \(\bigl( \goodtr{R}\comp R^*\bigr)^+ \subseteq W\), and also \( (\goodtr{R})^+ \subseteq W\).

\begin{lemma}\label{lem:periodic}
  Every infinite \(R\)-trace has a suffix that is an infinite \(\badtr{R}\)-trace.
\end{lemma}
\begin{proof} 
  Assume the contrary, i.e., there exists an infinite \(R\)-trace
  $s_1,s_2,\ldots$ that contains infinitely many steps from
  \(\goodtr{R}\).
  Let $S=s_{i_1},s_{i_2},\ldots$ be the infinite subsequence of states
  such that \((s_{i_j},s_{i_j+1}) \in \goodtr{R}\) for all \(j\geq 1\).
  Recall also that \(W=W_1\cup\cdots\cup W_n\) where each \(W_{\ell}\) is
  well-founded.
  For any \( s_i,s_j \in S \) with $i<j$ it holds that \((s_i,s_j)\in
  \goodtr{R}\comp R^* \), and thus, according to
  Lem.~\ref{lem:goodtrwff}, we also have that \( (s_i,s_j)\in W_{\ell} \)
  for some \( 1\le \ell \le n\).
  Ramsey's theorem~\cite{Rams29} guarantees the existence of an infinite subsequence
  \( S'= s_{j_1},s_{j_2},\ldots\) of \(S\), and a single \(W_{\ell}\), such
  that for all \( s_i,s_j \in S' \) with $i<j$ we have \( (s_i,s_j)\in W_{\ell}
  \). This contradicts that \(W_{\ell}\) is well-founded and we are done.\qed
\end{proof}

\begin{remark}
  When fixpoints are not computable, they can be approximated from above or
  from below \cite{Cousot77-POPL}.  It is routine to check that the results of
  Lemmas~\ref{lem:goodtrwff} and \ref{lem:periodic} remain valid when replacing
  \(G=\gfp\ \lambda Y\ldotp W\cap \tilde{g}(Y)\) in Def.~\ref{def:partition} with \(G' \subseteq
  \gfp\ \lambda Y\ldotp W\cap \tilde{g}(Y)\).  Therefore we have that, even when
  approximating \(\gfp\ \lambda Y\ldotp W\cap \tilde{g}(Y) \) from below, the
  termination guarantees of \(\{\goodtr{R},\badtr{R}\}\) still hold. In
  Sec.~\ref{sec:implementation}, we shall see how to exploit this result in
  practice.
\end{remark}

%

\begin{example}[cont'd from Ex.~\ref{ex:firstrb}]\label{ex:ndrb}
  We left Ex.~\ref{ex:firstrb} with \(W=\{x'<x, x\geq 1 \}\) and
  \(\badtr{R}=\{x\geq1,x'=x+y,y'=y-1,y\geq0\} \). 
  As argued previously, to prove the well-foundedness of \(R\) it is enough to
  show that \(\badtr{R}\) is well-founded. For clarity, we rename \(\badtr{R}\) into \(\badtr{R}^{(1)}\).
  Next we partition \(\badtr{R}^{(1)}\) as we did it for \(R\) in Ex.~\ref{ex:firstrb}.
  As a result, we update \(W\) by adding the
  well-founded relation \( \{y'<y, y\geq 0\} \).  Then we evaluate
  again \(G\) (we omit calculations) which yields \(\badtr{R}^{(2)}=\emptyset\). 
  Hence we conclude from Lem.~\ref{lem:periodic} that \(R\) is well-founded.\hfill\(\blacksquare\)
\end{example}

Building upon all the previous results, we introduce \acabar that is given at Alg.~\ref{alg:acabar}. \texttt{Acabar} is a recursive procedure that takes as input two
parameters: a transition relation \(R\) and a disjunctively well-founded relation \(W\).
The second parameter is intended for recursive calls, hence the user should
invoke \texttt{Acabar} as follows: {\ttfamily Acabar(\(R,\emptyset\))}.
We call it the \emph{root call}.
Upon termination, \texttt{Acabar} returns a subset \(\badtr{R}\) of the
transition relation \(R\).  If it returns the empty set, then the relation \(R\) is well-founded, hence termination is proven.
Otherwise (\(\badtr{R}\neq \emptyset\)), we can not know for sure if
\(R\) is well-founded: there might be an infinite \(R\)-trace. However, 
Lem.~\ref{lem:periodic} tells us that every infinite \(R\)-trace must have a suffix that is an infinite \(\badtr{R}\)-trace.
It may also be the case that \(\badtr{R}\) is well-founded (and so is \(R\)) in
which case it was not discovered by \texttt{Acabar}.  Another case is that
\(R=\badtr{R}\). In this case we have made no progress and therefore we stop.
Whenever \(\badtr{R}\neq\emptyset\), we call this returned value the
\emph{problematic} subset of \(R\).


Next we study progress properties of \acabar. We start by defining the sequence
\(\{R^{(i)}\}_{i\geq 0}\) where each \(R^{(i)}\) is the argument passed to the
\(i\)-th recursive call to \acabar. 
In particular, \(R^{(0)}\) is the argument of the root call. Furthermore, we
define the sequences \(\{\badtr{R}^{(i)}\}_{i\geq 1}\) and
\(\{\goodtr{R}^{(i)}\}_{i\geq 1}\) where 
\(\{\goodtr{R}^{(i)},\badtr{R}^{(i)}\}\) is a partition of \(R^{(i-1)}\) and \(\badtr{R}^{(i)}=R^{(i)}\) for all \(i\geq 1\).

\begin{lemma}\label{lem:progress}
 Let a run of \acabar with at least \(i \geq 1\) recursive calls,
  then we have\\
\hspace*{\stretch{1}}\( R^{(0)} \supsetneq  R^{(1)} \supsetneq\cdots \supsetneq 
   R^{(i)} \enspace .\)\hfill\vspace{0pt}
\end{lemma}
\begin{proof}
  The proof is by induction on \(i\), for \(i=1\) it follows from the
  definitions that \(R^{(1)}=\badtr{R}^{(1)}\) and
  \(\{\badtr{R}^{(1)},\goodtr{R}^{(1)}\}\) is a partition of \(R^{(0)}\).
  Moreover, since at least \(i=1\) recursive calls take place we find that
  the condition of line~\ref{ln:conditional} fails, meaning neither
  \(\badtr{R}^{(1)}\) nor \(\goodtr{R}^{(1)}\) is empty, hence
  \(R^{(1)}\) is a strict subset of \(R^{(0)}\). The inductive case is similar.\qed
\end{proof}

By Lemmas~\ref{lem:periodic} and~\ref{lem:progress}, we have that
every infinite \(R^{(0)}\)-trace has a suffix that is an infinite
\(\badtr{R}^{(i)}\)-trace for every \(i\geq 1\). As a consequence,
forcing \acabar to execute line~\ref{ln:ret} after predefined number
of recursive calls, it returns a relation \(\badtr{R}^{(i)}\) such that the previous
property holds. Incidentally, we find that \acabar proves program termination when it returns the empty set as stated next.

\begin{algorithm}[t]\label{alg:acabar}
\caption{Enhanced modular reasoning}
	\DontPrintSemicolon
	\LinesNumbered
	\SetKwFunction{acabar}{Acabar}
	\SetKwFunction{computew}{find\_dwf\_candidate}
	\acabar{R,W}\;
	\KwIn{a relation \(R\subseteq \mathcal{Q}\times \mathcal{Q} \)}
	\KwIn{a relation \(W\subseteq \mathcal{Q}\times \mathcal{Q} \) such that \(W\) is disjunctively well-founded}
	\KwOut{ \( \badtr{R} \subseteq R\) }
	\Begin{
	\(W\coloneqq W\cup\)\computew{\( R \)} \label{ln:computeti}\;
	   let \(G\) be such that \( G \subseteq \gfp\ \lambda Y\ldotp W\cap \tilde{g}(Y) \)\label{ln:computeG} \;
	    \( \badtr{R} \coloneqq  R \setminus G \) \label{ln:computerb}\;
	    \eIf{ \( \badtr{R} = \emptyset \) or \( \badtr{R}=R \)\label{ln:conditional}}{ 
	    \Return \( \badtr{R} \) \label{ln:ret}\;
	    }{ 
	        \Return \acabar{\( \badtr{R}, W\)}\;
	    }
	}
\end{algorithm}

\begin{theorem}
  Upon termination of the call {\ttfamily Acabar(\(R,\emptyset\))}, if it
  returns the empty set, then the relation \(R\) is well-founded.
  \label{th:acabar}
\end{theorem}

Let us now turn to line~\ref{ln:computeti}. There, \acabar
calls a subroutine {\ttfamily find\_dwf\_candidate(\(R\))}
implementing a heuristic search which returns a disjunctively well-founded
relation using hints from the representation and the domain of \(R\). 
Details about its implementation, that is inspired from previous work
\cite{CookGLRS08,CookPR05}, will be given at Sec.~\ref{sec:experiments} 
 --- we will consider the case of \(R\) being a relation over the integers of the form
\(R=\rho_1\vee\cdots\vee\rho_n\) where each \(\rho_i\) is a conjunction of
linear constraints over the variables \(\bar{x}\) and \(\bar{x}'\).
Let us intuitvely explain this procedure on an example.


\begin{example}[cont'd from Ex.~\ref{ex:ndrb}]
  {\ttfamily Acabar(\(R,\emptyset\))} updates \(W\) as follows: 
\begin{inparaenum}[\upshape(\itshape 1\upshape)]
\item \(\emptyset\);
\item \(\{x'<x, x\geq 1\}\);
\item \(\{x'<x, x\geq 1\}, \{y'<y, y\geq 0\}\).
\end{inparaenum}
  The first update from \(\emptyset\) to \(\{x'<x, x\geq
  1\}\) is the result of calling {\ttfamily find\_dwf\_candidate(\(R\))}. 
  The hint used by {\ttfamily find\_dwf\_candidate} is that \(x\) is bounded
  from below in \(R\).
  The second update to \(W\) results from calling {\ttfamily
  find\_dwf\_candidate(\(\badtr{R}=\{x\geq1,x'=x+y,y'=y-1,y\geq0\}\))}.
  Since \(\badtr{R}\) has the linear ranking function \(f(x,y)=y\), 
  {\ttfamily find\_dwf\_candidate} returns \(\{y'<y, y\geq 0\}\).\hfill\(\blacksquare\)
\end{example}

%

\section{\texttt{Acabar} for Conditional Termination}
\label{sec:cond}

As mentioned previously, upon termination, \texttt{Acabar} returns a subset
\(\badtr{R}\) of the transition relation \(R\). If this set is empty then \(R\)
is well-founded and we are done. Otherwise, \(\badtr{R}\) is a non-empty subset
and called the problematic set.  In this section, we shall see how to compute,
given the problematic set, a \emph{precondition \(\mathcal{P}\) for termination}.  More
precisely, \(\mathcal{P}\) is a set of states such that no infinite \(R\)-trace starts with
a state of \(\mathcal{P}\).
We illustrate our definitions using the simple but challenging example of Sec.~\ref{sec:motivating}.

\begin{example}\label{ex:ct:1}
  Consider again the relation %
%
%
\(R=\{x>0,x'=x+y,y'=y+z, z'=z\}\).
Upon termination \texttt{Acabar} returns the following relation:
\[
\badtr{R}=\{x'=x+y,y'=y+z,z'=z,x>0,y\geq 0,z\geq 0\}
\]
which corresponds to all the cases where \(x\) is stable or increasing
over time.\hfill\(\blacksquare\)
\end{example}

Lemma~\ref{lem:periodic} tells us that every infinite \(R\)-trace \(\pi\)
is such that \(\pi=\pi_f \pi_{\infty} \) where \(\pi_f\) is a finite
\(R\)-trace and \(\pi_{\infty}\) is an infinite \(\badtr{R}\)-trace.
%
%
Our computation of a precondition for termination is divided into the following parts: 
\begin{inparaenum}[\upshape(\itshape i\upshape)]
\item compute those states \(\mathcal{Z}\) visited by infinite \(\badtr{R}\)-trace;
\item compute the set \(\mathcal{V}\) of \(R^*\)-predecessors of
  \(\mathcal{Z}\), that is the set of states visited by some \(R\)-trace ending in \(\mathcal{Z}\); and
\item compute \(\mathcal{P}\) as the complement of \(\mathcal{V}\).
\end{inparaenum}
Formally, \upshape(\itshape i\upshape) is given by a greatest fixpoint
expression \(\gfp\ \lambda \mathcal{X}\ldotp \pre[\badtr{R}](\mathcal{X})\).
This expression is directly inspired by the work of Bozga et al.~\cite{BozgaIK12} on deciding conditional termination. This
greatest fixpoint is the largest set \(\mathcal{Z}\) of states each
of which has an \(\badtr{R}\)-successor in \(\mathcal{Z}\).
Because of this property, every  infinite \(\badtr{R}\)-trace visits only 
states in \(\mathcal{Z}\).
%
In \(\pi=\pi_f\pi_{\infty}\), this corresponds to the suffix \(\pi_{\infty}\)
that is an infinite \(\badtr{R}\)-trace.

\begin{example}\label{ex:ct:2}
For \(\badtr{R}\) as given in Ex.~\ref{ex:ct:1}, we have that 
\(
\mathcal{Z}=\{ z\geq 0,y\geq 0,x>0 \}
\)
which contains the following infinite \(\badtr{R}\)-trace:\\
\hspace*{\stretch{1}}
\(
(x=1,y=0,z=0) \mathbin{\badtr{R}} (x=1,y=0,z=0) \mathbin{\badtr{R}} (x=1,y=0,z=0) \mathbin{\badtr{R}} \ldots 
\)%
\hfill\(\blacksquare\)
\end{example}

Let us now turn to  \upshape(\itshape ii\upshape), that is computing the set
\(\mathcal{V}\) of \(R^*\)-predecessors of \(\mathcal{Z}\). It is known that
\(\mathcal{V}\) coincides with \(\lfp\ \lambda \mathcal{X}\ldotp
\mathcal{Z}\cup \pre[R](\mathcal{X})\).  Intuitively, we prepend to those
infinite \(\badtr{R}\)-traces a finite \(R\)-trace. That is,
prefixing \(\pi_f\) to \(\pi_{\infty}\) results in \(\pi=\pi_f\pi_{\infty}\).
Finally, step \upshape(\itshape iii\upshape) results into a
 precondition for termination \(\mathcal{P}\) obtained by complementing
 \(\mathcal{V}\).

\begin{example}
Computing \(\lfp\ \lambda \mathcal{X}\ldotp \mathcal{Z}\cup \pre[R](\mathcal{X})\) for 
\(\mathcal{Z}\) as given in Ex.~\ref{ex:ct:2} and \linebreak \(R=\{x'=x+y,y'=y+z,z'=z,x>0,y\geq 0,z\geq 0\}\) (Ex.~\ref{ex:ct:1}) gives 
$\mathcal{V}=\mathcal{V}_1\lor\mathcal{V}_2$ where
\begin{align*}
\mathcal{V}_1&= \{x\geq 1,z=0,y\geq 0\}\\
\mathcal{V}_2&= \{x\geq 1,z\geq 1\}\cup\{x+i*y+j*z \geq 1 \mid i \ge 1, j=\textstyle{\sum_{k=0}^{i-1}k}\}\enspace .
\end{align*}
Intuitively, the set $\mathcal{V}_1$ of states corresponds to entering the loop with $z=0$ and $y$
non-negative, in which case the loop clearly does not terminate.
The set $\mathcal{V}_2$ of states corresponds to entering the loop with
$z$ positive, and the loop does not terminate after $i$-th iterations
for all $i$. Note that $\mathcal{V}_2$ consists of infinitely many
atomic formulas.
Complementing \(\mathcal{V}\) gives \(\mathcal{P}\).\hfill\(\blacksquare\)
%
\end{example}

\begin{theorem}
  There exists an infinite \(R\)-trace starting from \(s\) if{}f \(s\notin\mathcal{P}\).
\end{theorem}

\smallskip \noindent %
{\it Approximations.} %
As argued previously, it is often the case that only approximations of
fixpoints are available. In our case, any overapproximation of either
\(\mathcal{Z}\) or \(\mathcal{V}\) can be exploited to infer \(\mathcal{P}\).
Because of approximations, we lose the if direction of the theorem, that is,
we can only say that there is no infinite \(R\)-trace starting from some
\(s\in\mathcal{P}\).

\begin{example}
Using finite disjunctions of linear constraints, we can approximate
  \(\mathcal{V}\) by
\[
\{x\geq 1,z=0,y\geq 0\} \lor
\{x\geq 1,z\geq 1,x+y\geq 1,x+2y+z\geq 1,x+3y+3z\geq 1\}
\]
and then the complement \(\mathcal{P}\) is
\[
x\leq 0 \lor
x+y\geq 1 \lor x+2y+z\geq 1 \lor x+3y+3z\geq 1
\lor z\leq  -1 \lor (y\leq  -1 \land z\leq 0)
\]
which is a sufficient precondition for termination. Note that the
first $4$ disjuncts correspond to the executions which terminates after $0$, $1$, $2$ and $3$ iterations.\hfill\(\blacksquare\)
\end{example}

%

\section{Implementation}\label{sec:implementation}

We have implemented the techniques described in Sec.~\ref{sec:modular} and
\ref{sec:cond} for the case of multiple-path integer linear-constraint loops.
These loops correspond to relations of the form
\(R=\rho_1\vee\cdots\vee\rho_d\) where each \(\rho_i\) is a
conjunction of linear constraints over the variables \(\bar{x}\) and
\(\bar{x}'\). In this context, the set $\mathcal{Q}$ of states is
equal to $\mathbb{Z}^n$ where \(n\) is the number of variables in \(\bar{x}\).
This is a classical setting for
termination~\cite{Ben-AmramG13,BradleyMS05,PR04}.
Internally, we represent sets of states and relations over them as DNF
formulas where the atoms are linear constraints.
In what follows, we explain sufficient implementation details so that our
experiments can be independently reproduced if desired. Our
implementation is available~\cite{acabar}.

We start with line~\ref{ln:computeti} of Alg.~\ref{alg:acabar}. Recall
that the purpose of this line is to add more well-founded relations to
\(W\) based on the current relation \(R\).
In our implementation, \(W\) consists of well-founded relations of the form
\(\{f(\bar{x}) \ge 0, f(\bar{x}')<f(\bar{x})\}\) where \(f\) is a
linear function~\cite{CookPR05,CookGLRS08}.
Thus, our 
implementation looks for such well-founded
relations.
%
In particular, for each \(\rho_i\) of \(R\) we
add new well-founded relations to \(W\) as follows:
if \(\rho_i\) has a linear ranking function \(f(\bar{x})\) that is
synthesized automatically~\cite{PR04,Ben-AmramG13} then
\(\{f(\bar{x}')<f(\bar{x}), f(\bar{x})\ge 0 \}\) is added to \(W\);
otherwise,
let \(\{f_1(\bar{x})\ge 0, \ldots, f_d(\bar{x})\geq 0\}\) be the
result of projecting each \(\rho_i\) on \(\bar{x}\) (i.e., eliminating
variables \(\bar{x}'\) from \(\rho_i\)), then \(\{ 
\{f_i(\bar{x}')<f_i(\bar{x}), f_i(\bar{x}) \geq 0\} \mid 1\leq i \leq d \}\) is
added to \(W\). 
Because \(f_i\) is bounded but not necessarily decreasing, it
is called a \emph{potential linear ranking function}~\cite{CookGLRS08}.

As for line~\ref{ln:computeG}, recall that
\(G\) is a subset of \(\gfp\ \lambda Y\ldotp W\cap \tilde{g}(Y)\).
Furthermore, the sole purpose of \(G\) is to compute \(\badtr{R}=R\setminus
G\).  We now observe that \(\neg G\), the complement of \(G\), is as good as
\(G\).  In fact, \(\badtr{R}=R\cap (\neg G)\).
So by considering \(\neg G\) instead, what we are looking for is an
overapproximation of \(\neg (\gfp\ \lambda Y\ldotp W\cap \tilde{g}(Y))\).
Next we recall Park's theorem replacing the above
expression by a least fixpoint expression.

\begin{theorem}[From~\cite{park69}]\label{thm:park}
Let \(\tuple{L,\sqsubseteq,\bigsqcap,\bigsqcup,\top,\bot,\neg}\) be a complete
Boolean algebra and let \(f\in L\rightarrow L\) be an order-preserving function then
\(f'=\lambda X\ldotp \neg(f(\neg X))\) is an order-preserving function on \(L\) and
\(\neg\bigl(\gfp\ f\bigr)=\lfp\ f'\).
\end{theorem}

Park's theorem applies in our setting because computations are carried over
the Boolean algebra \(\tuple{2^{(\mathcal{Q}\times\mathcal{Q})}, \subseteq, \cap,\cup, (\mathcal{Q}\times\mathcal{Q}), \emptyset, \neg}\).
Applying it to \(\gfp\ \lambda Y\ldotp W\cap \tilde{g}(Y)\) where
\(\tilde{g}(Y)=\neg(\neg Y\comp R^{-1})\), we find that 
\[ 
\neg\bigl(\gfp\ \lambda Y\ldotp W\cap
\neg(\neg Y\comp R^{-1})\bigr) = \lfp\ \lambda Y\ldotp (\neg W)\cup Y\comp R^{-1}\enspace .
\]
Therefore, to implement line~\ref{ln:computeG}, we rely on abstract
interpretation to compute an overapproximation of \(\lfp\ \lambda Y\ldotp (\neg
W)\cup Y\comp R^{-1}\), hence, by negation, an underapproximation of \(\gfp\
\lambda Y\ldotp W\cap \tilde{g}(Y)\) therefore complying with the requirement
on \(G\).

As far as abstract interpretation is concerned, our implementation uses a
combination of predicate abstraction~\cite{GS97} and trace
partitioning~\cite{MauborgneR05}. 
The set of predicates is given by a finite set of atomic linear
constraints and is also closed under negation, e.g., if $x+y \ge 0$
is a predicate then $x+y \le -1$ is also a predicate.
Abstract values are positive Boolean combination of atoms taken from the set of
predicates.  Observe that although negation is forbidden in the definition of
abstract values, the abstract domain is closed under complement.


The set of predicates is chosen so as the following invariant to hold: each
time the control hits line~\ref{ln:computeG}, the set contains enough
predicates to represent precisely each well-founded relation in \(W\).
Our implementation provides enhanced precision by enforcing a stronger
invariant: besides the above predicates for \(W\), it includes all
atomic linear constraints occurring in the formulas representing
\(X_1,\ldots,X_\ell\) where \(\ell\geq0\), \(X_0 =(\neg W)\) and
\(X_{i+1}= (\neg W)\cup X_i \comp R^{-1}\). The value of \(\ell\) is
user-defined and, in our experiments, it did not exceed $1$. %

To further enhance precision at line~\ref{ln:computeG}, we apply trace
partitioning~\cite{MauborgneR05}. The set of \(R\)-traces is
partitioned using the linear atomic constraints of the form
\(f(\bar{x}')<f(\bar{x})\) that appear in \(W\).  
More precisely, partitioning \(R\) on \(f(\bar{x}')<f(\bar{x})\) is
done by replacing each \(\rho_i\)  by \((\rho_i\land
f(\bar{x}')<f(\bar{x})) \lor (\rho_i\land f(\bar{x}')\geq f(\bar{x}))\).

As for conditional termination, overapproximating \(\mathcal{Z} =
\gfp\ \lambda \mathcal{X}\ldotp \pre[\badtr{R}](\mathcal{X}) \) is done by computing
the last element \(\mathcal{X}_{\ell}\) from the finite sequence
\(\mathcal{X}_0,\ldots,\mathcal{X}_{\ell}\) given by \(\mathcal{X}_0
=\mathcal{Q}\) and \(\mathcal{X}_{i+1}= \mathcal{X}_{i}\land\pre[\badtr{R}](\mathcal{X}_i)\) 
where
\(\ell\) is predefined. The result is always representable as DNF formula where
the atoms can be any atomic linear constraints.
As for \(\mathcal{V} = \lfp\ \lambda \mathcal{X}\ldotp \mathcal{X}_\ell\cup
\pre[R](\mathcal{X})\), an overapproximation is computed in a similar
way to that of line~\ref{ln:computeG}, i.e.,
using a combination of predicate abstraction and trace partitioning.

%

\section{Experiments}\label{sec:experiments}

%

\begin{table}

\begin{tabular}{|r||l|l|}

\hline
\(\sharp\) & loop & termination precondition \\
\hline\hline

1
&
\begin{lstlisting}
while (x$\geq$0) x'=-2x+10;
\end{lstlisting}
&
\anyinput
\\ \hline

2
&
\begin{lstlisting}
while (x>0) x'=x+y; y'=y+z;
\end{lstlisting}
& 
\begin{minipage}{3.3cm}
\(x {\leq} 0\lor z{<}0 \lor\) \\
\((z{=}0 \land y{<}0) \lor\)\\
\(x{+}y{\leq} 0 \lor x{+}2y{+}z{\leq} 0 \lor \)\\
\(x{+}3y{+}3z {\leq} 0\)
\end{minipage}
\\ \hline

3
&
\begin{lstlisting}
while (x$\leq$N)
  if (*) { x'=2*x+y;  y'=y+1; } else x'=x+1;
\end{lstlisting}
&
\( x>n \lor x+y \ge 0 \)
\\ \hline

4
&
\begin{lstlisting}
@requires n>200 and y<9
while (1)
  if (x<n) {
    x'=x+y;
    if (x'$\geq$200) break;
  } 
\end{lstlisting}
&
\begin{minipage}{3.3cm}
 \( n\le 200 \lor  y\ge 9 \lor \) \\
 \( (x<n \land y\ge 1) \lor \) \\
 \( (x{<}n \land x{\ge} 200\land x{+}y{\ge}200) \)
\end{minipage}\\
\hline

5
&
\begin{lstlisting}
while (x<>y) if (x>y) x'=x-y; else y'=y-x;
\end{lstlisting}
\hfill\optimal
&
\( (x \ge 1 \land y\geq 1) \lor x=y \)
\\ 

\hline\hline

6
&
\begin{lstlisting}
while (x<0) x'=x+y; y'=y-1;
\end{lstlisting}
&
\begin{minipage}{3.3cm}
\( x \ge 0 \lor x+y \ge 0 \lor \\
x+2y\geq 1 \lor x+3y \geq 3\)
\end{minipage}
\\ \hline

7
&
\begin{lstlisting}
while (x>0) x'=x+y; y'=-2y;
\end{lstlisting}
\hfill\optimal
&
\( x \leq 0 \lor y\neq 0 \)
\\ \hline

8
&
\begin{lstlisting}
while (x<y) x'=x+y; y'=-2y;
\end{lstlisting}
\hfill\optimal
&
\( x \geq 0 \lor y\neq 0 \)
\\ \hline

9
&
\begin{lstlisting}
while (x<y) x'=x+y; 2y'=y;
\end{lstlisting}
\hfill\optimal
&
\( x \geq 0 \lor y\neq 0 \)
\\ \hline

10
&
\begin{lstlisting}
while (4x-5y>0) x'=2x+4y; y'=4x;
\end{lstlisting}
\hfill\(\optimal\)
&
\begin{minipage}{3.5cm}
\( 5y-4x \geq 0 \vee \\
   (3x-4y \geq 0 \wedge 16x-21y \geq 1)
\)
\end{minipage}
\\ \hline

11
&
\begin{lstlisting}
while (x<5) x'=x-y; y'=x+y;
\end{lstlisting}
\hfill\optimal
&
\( x\neq 0 \lor y \neq 0 \)
\\ \hline

12
&
\begin{lstlisting}
while (x>0 and y>0) x'=-2x+10y;
\end{lstlisting}
\hfill\(\optimal\)
&
\(x \leq 3 \lor 10y-3x \neq 0\)
\\ \hline

13
&
\begin{lstlisting}
while (x>0) x'=x+y;
\end{lstlisting}
&
\(x\le 0 \lor y<0 \lor x+y \le 0\)
\\ \hline

14
&
\begin{lstlisting}
while (x<10) x'=-y; y'=y+1;
\end{lstlisting}
\hfill\optimal
&
\( y\leq -10 \lor x \geq 10 \)
\\ \hline

15
&
\begin{lstlisting}
while (x<0) x'=x+z; y'=y+1; z'=-2y
\end{lstlisting}
&
\( x \geq 0 \lor x+z\ge 0 \)
\\

\hline\hline

16
&
\begin{lstlisting}
while (x>0 and x<100) x'$\geq$2x+10;
\end{lstlisting}
\hfill\linear
&
\anyinput
\\ \hline

17
&
\begin{lstlisting}
while (x>1) -2x'=x;
\end{lstlisting}
\hfill\linear
&
\anyinput
\\ \hline

18
&
\begin{lstlisting}
while (x>1) 2x'$\leq$x;
\end{lstlisting}
\hfill\linear
&
\anyinput
\\ \hline

19
&
\begin{lstlisting}
while (x>0) 2x'$\leq$x;
\end{lstlisting}
\hfill\linear
&
\anyinput
\\ \hline

20
&
\begin{lstlisting}
while (x>0) x'=x+y; y'=y-1;
\end{lstlisting}
&
\anyinput
\\ \hline

21
&
\begin{lstlisting}
while (4x+y>0) x'=-2x+4y; y'=4x;
\end{lstlisting}
&
\begin{minipage}{3.5cm}
\( 4x+y \leq 0 \vee \\
 ( x-4x\geq 0 \wedge 8x-15y \geq 1) \)
\end{minipage}
\\ \hline

22
&
\begin{lstlisting}
while (x>0 and x<y) x'=2x; y'=y+1;
\end{lstlisting}
&
\anyinput
\\ \hline

23
&
\begin{lstlisting}
while (x>0) x'=x-2y; y'=y+1;
\end{lstlisting}
&
\anyinput
\\ \hline

24
&
\begin{lstlisting}
while (x>0 and x<n) x'=-x+y-5; y'=2y; n'=n;
\end{lstlisting}
&
\anyinput
\\ \hline

25
&
\begin{lstlisting}
while (x>0 and y<0) x'=x+y; y'=y-1;
\end{lstlisting}
\hfill\linear
&
\anyinput
\\ \hline

26
&
\begin{lstlisting}
while (x-y>0) x'=-x+y; y'=y+1;
\end{lstlisting}
&
\anyinput
\\ \hline

27
&
\begin{lstlisting}
while (x>0) x'=y; y'=y-1;
\end{lstlisting}
&
\anyinput
\\ \hline

28
&
\begin{lstlisting}
while (x>0) x'=x+y-5; y'=-2y;
\end{lstlisting}
&
\anyinput
\\ \hline

29
&
\begin{lstlisting}
while (x+y>0) x'=x-1; y'=-2y;
\end{lstlisting}
&
\anyinput
\\ \hline

30
&
\begin{lstlisting}
while (x>y) x'=x-y; 1$\leq$y'$\leq$2
\end{lstlisting}
\hfill\linear
&
\anyinput
\\ \hline

31
&
\begin{lstlisting}
while (x>0) x'=x+y; y'=-y-1;
\end{lstlisting}
&
\anyinput
\\ \hline

32
&
\begin{lstlisting}
while (x>0) x'=y; y'$\leq$-y;
\end{lstlisting}
\hfill\linear
&
\anyinput
\\ \hline

33
&
\begin{lstlisting}
while (x<y) x'=x+1; y'=z; z'=z;
\end{lstlisting}
&
\anyinput
\\ \hline

34
&
\begin{lstlisting}
while (x>0) x'=x+y; y'=y+z; z'=z-1;
\end{lstlisting}
&
\anyinput
\\ \hline

35
&
\begin{lstlisting}
while (x+y$\geq$0 and x$\leq$z) x'=2x+y; y'=y+1; z'=z
\end{lstlisting}
&
\anyinput
\\ \hline

36
&
\begin{lstlisting}
while (x>0 and x$\leq$z) x'=2x+y; y'=y+1; z'=z
\end{lstlisting}
&
\anyinput
\\ \hline

37
&
\begin{lstlisting}
while (x$\geq$0) x'=x+y; y'=z; z'=-z-1;
\end{lstlisting}
&
\anyinput
\\ \hline

38
&
\begin{lstlisting}
while (x-y>0) x'=-x+y; y'=z; z'=z+1;
\end{lstlisting}
&
\anyinput
\\ \hline

39
&
\begin{lstlisting}
while (x>0 and x<y) x'>2x; y'=z; z'=z;
\end{lstlisting}
&
\anyinput
\\ \hline

40
&
\begin{lstlisting}
while (x$\geq$0 and x+y$\geq$0) x'=x+y+z; y'=-z-1; z'=z;
\end{lstlisting}
\hfill\linear
&
\anyinput
\\ \hline

41
&
\begin{lstlisting}
while (x+y$\geq$0 and x$\leq$n) x'=2x+y; y'=z; z'=z+1; n'=n;
\end{lstlisting}
&
\anyinput
\\ \hline
\end{tabular}
\caption{Benchmarks used in experiments. Loops (1--5) are taken from~\cite{CookGLRS08} and (6--41) from \cite{ChenFM12}.}
\label{tb:bench}
\end{table}

We have evaluated our prototype implementation against a set of benchmarks
collected from publications in the area~\cite{CookGLRS08,ChenFM12}.
In what follows, we present the results of our implementation for those loops,
and compare them to existing tools for proving
termination~\cite{armc,ChenFM12,BradleyMS05} as well as tools for inferring
preconditions for termination~\cite{CookGLRS08}.  We compare the different
techniques according to what the corresponding implementations report. We
ignore performance because, for the selected benchmarks, little insight can be
gained from performance measurements when an implementation was available
(which was not always the case \cite{rybal-perso}).

The benchmarks accompanied with our results are depicted in Table~\ref{tb:bench}.
Translating each loop to a relation of the form \(R=\rho_1\vee\cdots\vee\rho_n\) is straightforward.
Every line in the table includes a loop and its inferred termination
precondition (\anyinput means it terminates for any input).
In addition, preconditions (different from \anyinput) marked with
\optimal are optimal, i.e., the corresponding loop is non-terminating
for any state in the complement.

We have divided the benchmarks into \(3\) groups: (\(1\)--\(5\)), (\(6\)--\(15\)) and (\(16\)--\(41\)).
With the exception of loop \(1\), each loop in group (\(1\)--\(5\)) includes non-terminating
executions and thus those loops are suitable for inferring preconditions.
Our implementation reports the same preconditions as the tool of Cook et
al.~\cite{CookGLRS08} 
save for loop 1 for which their tool is reported to infer the
precondition \( x>5 \vee x<0 \), while we prove termination for all input. Note
that every other tool used in the comparison \cite{ChenFM12,BradleyMS05,armc}
fail to prove termination of this loop. 
Further, the precondition we infer for loop \(5\) is optimal.

All the loops (\(6\)--\(15\))  are non-terminating.  Chen et al.~\cite{ChenFM12}
report that their tool cannot handle them since it aims at proving termination
and not inferring preconditions for termination.
We infer preconditions for all of them, and in addition, most of them
are optimal (those marked with \(\bullet\)).
Unfortunately for those loops we could not compare with the tool of Cook et
al.~\cite{CookGLRS08}, since there is no implementation
available~\cite{rybal-perso}.

Loops in the group (\(16\)--\(41\)) are all terminating. 
Those marked with \linear{} actually have linear
ranking functions, those unmarked require disjunctive well-founded
transition invariants with more than one disjunct.
We prove termination of all of them except loop \(21\). We point that the tool of
Chen et al.~\cite{ChenFM12} also fails to prove termination of loop 21, but
also of loop \(34\). On the other benchmarks, they prove termination.  They also
report that PolyRank~\cite{BradleyMS05} failed to prove termination of any of
the loops that do not have a linear ranking function.
In addition, we applied ARMC~\cite{armc} on the loops of the group (\(16\)--\(41\)).
ARMC, a transition invariants based prover, succeeded to prove
termination for all those loops with a linear ranking function (marked with
\linear) and also loop \(39\).

Next we discuss in details the analysis of two selected examples from
Table~\ref{tb:bench}.

\begin{example}
\label{ex:exp:1}
Let us explain the analysis of loop \(1\) in details starting 
with the root call \(\acabar(R,\emptyset)\) where \(R=\{x\geq0,
x'=-2x+10\}\).
At line~\ref{ln:computeti}, since \(R\) includes the bound \(x\geq0\),
i.e., \(f(x)=x\) is a potential linear ranking function, we add \( \{x'<x,
x\geq 0\} \) to \(W\).
Computing \(G\) at line~\ref{ln:computeG}, hence \(\badtr{R}\) at
the following line, results in \(\badtr{R}=\rho_1\lor\rho_2\) where
\(\rho_1=\{x'= -2x+10, x\geq 0, x \leq 3\}\) and \(\rho_2= \{x'=
-2x+10, x\geq 4, x \leq 5\}\).

Note that \(\rho_1\) is enabled for \(0\leq x \leq 3\) and in this case
\(x'>x\).  Also \(\rho_2\) is enabled for \(x=4\) or \(x=5\) for which \(x'<x\)
and thus \(\rho_2\subseteq W\), however, after one more iteration, the value of
\(x\) increases (this is why \(\rho_2\) is included in \(\badtr{R}\)).
Transitions for which \(x>5\) are not included in \(\badtr{R}\), hence they
belong to \(\goodtr{R}\) itself included in \(W\) (Lem.~\ref{lem:goodtrwff}).
Hence when \(x>5\) termination is guaranteed, this is also easily seen since
those transitions terminate after one iteration.

Since \(\badtr{R}\) is neither empty nor equal to \(R\), a recursive
call to \(\acabar(\badtr{R},W)\) takes place. 
At line~\ref{ln:computeti}, we add \(\{ -x' < -x, 10-x\geq 0\}\) to \(W\) since \(f(x)=10-x\) is
a linear ranking function for \(\rho_1\). Note that \(\rho_2\) has the linear
ranking function \(f(x)=x\) already included in \(W\).  Computing
\(G\) at line~\ref{ln:computeG}, hence \(\badtr{R}\), yields
\(\badtr{R}=\emptyset\) and therefore we conclude that the loop terminates for
any input.\hfill\(\blacksquare\)
\end{example}

\begin{example}
Let us explain the analysis of loop \(9\) in details
starting with the root call \(\acabar(R,\emptyset)\) where
\(R=\{x<y,x'=x+y, 2y'=y\}\).
At line~\ref{ln:computeti}, since \(R\) includes the bound \(y-x>0\),
i.e., \(f(x,y)=y-x-1\) is a potential linear ranking function, we add
\( \{y'-x'<y-x, y-x-1\geq 0\} \) to \(W\).
Computing \(G\) at line~\ref{ln:computeG}, hence \(\badtr{R}\) yields
\(\badtr{R}=\{x<y, x'=x+y, 2y'=y, y \leq 0\}\).
Note that \(\badtr{R}\) exclusively consists of transitions where $y$
is not positive, in which case $x'-y' \ge x-y$ and thus not included
in \(W\).
Transitions where $y$ is positive are not included in \(\badtr{R}\) (hence they
belong to \(\goodtr{R}\)) since they always decrease $x-y$, and thus
are transitively included in \(W\) (Lem.~\ref{lem:goodtrwff}).

Since \(\badtr{R}\) is neither empty nor equal to \(R\), we call
recursively \(\acabar(\badtr{R},W)\).  
At line~\ref{ln:computeti}, since \(R\) includes the bound \(y \leq
0\) (or equivalently \(-y \geq 0\)), i.e., \(f(x,y)=-y\) is a
potential linear ranking function, we add \( \{ -y'<-y, -y \geq 0\} \) to
\(W\).
Computing \(G\) at line~\ref{ln:computeG}, hence \(\badtr{R}\) yields
\(\badtr{R}=\{x<y, x'=x+y, 2y'=y, y = 0\}\).
Note that \(\badtr{R}\) exclusively consists of transitions where
$y=0$, which keeps both values of $x$ and $y$
unchanged. Transitions in which $y$ is negative belong to \(\goodtr{R}\),
hence they are transitively covered by \(W\) (Lem.~\ref{lem:goodtrwff}), in
particular by the last update (viz. \( \{ -y'<-y, -y \geq 0\} \)) to \(W\).

Since \(\badtr{R}\) is neither empty nor equal to \(R\), we call
recursively \(\acabar(\badtr{R},W)\). This time our implementation does not
further enrich \(W\) with a well-founded relation, and as a consequence,
after computing \(G\) at line~\ref{ln:computeG}, we get that
$\badtr{R}=R$. Hence, \acabar returns with 
\(\badtr{R}=\{x<y, x'=x+y, 2y'=y, y = 0\}\).

Now, given $\badtr{R}$, we infer a precondition for termination as described in
Sec.~\ref{sec:cond}. We first compute \(\gfp\ \lambda \mathcal{X}\ldotp
\pre[\badtr{R}](\mathcal{X})\), which in this case, converges in two steps with
$\mathcal{Z}\equiv y=0 \wedge x<0$.  Then we compute \(\lfp\ \lambda
\mathcal{X}\ldotp \mathcal{Z}\cup \pre[R](\mathcal{X})\), which results in
\(\mathcal{V}\equiv y=0 \land x<0\). The complement, \(\mathcal{P}\equiv y<0
\vee y>0 \vee x<0\), is a precondition for termination.
Note that the result is optimal, i.e., \(\mathcal{V}\) is a
precondition for non-termination. Optimality is achieved because
$\mathcal{Z}$ and $\mathcal{V}$ coincide with the $\gfp$ and the $\lfp$
of the corresponding operators, and are not overapproximations.\hfill\(\blacksquare\)
\end{example}


%
%

\section{Conclusion}
\label{sec:conclusion}

This work started with the invited talk of A.~Podelski at ETAPS\(\:\)'11 who
remarked that the inclusion check \(R^+ \subseteq W\) is equivalently
formulated as a safety verification problem where states are made of pairs.
Back to late 2007, a PhD thesis~\cite{Gan07} proposed a new approach to the
safety verification problem in which the author shows how to leverage the
equivalent backward and forward formulations of the inclusion check.  Those two
events planted the seeds for the backward inclusion check \(R\subseteq W^-\),
and later \(\acabar\).

\smallskip \noindent  %
{\it Initial States.} For the sake of simplicity, we deliberately
excluded the initial states \(\mathcal{I}\) from the previous developments.
Next, we introduce two possible options to incorporate knowledge about the
initial states in our framework.  The first option consists in
replacing \(R\) by \(R'\) that is given by \(R\cap
(\mathit{Acc}\times\mathit{Acc})\) where \(\mathit{Acc}\) denotes (an
overapproximation of) the reachable states in the system.  Formally,
\(\mathit{Acc}\) is given by the least fixpoint \(\lfp\
\lambda\mathcal{X}\ldotp \mathcal{I}\cup \post[R](\mathcal{X})\).

The second option is inspired by the work of Cousot
\cite{cousotphdthesis} where he mixes backward and forward reasoning. We give here some
intuitions and preliminary development. Recall that the
greatest fixpoint \(\gfp\ \lambda Y\ldotp W\cap \tilde{g}(Y)\)
of line~\ref{ln:computeG} is best understood as the result of
removing all those pairs \( (s,s')\in W\) such that \( (s,s') \comp R^+
\nsubseteq W\). We observe that the knowledge about initial states is not used
in the greatest fixpoint. A way to incorporate that knowledge is 
to replace the greatest fixpoint expression by the following
one \(\gfp\ \lambda Y\ldotp (B\cap W)\cap \tilde{g}(Y)\) where \(B\) takes the reachable states into account.
In a future work, we will formally develop those two options and evaluate their benefit. 

\smallskip \noindent %
{\it Related Works.} 
As for termination, our work is mostly related
to the work of Cook et al.~\cite{CookPR05,CookPR06} where the inclusion check \(R^+ \subseteq
W\)~\cite{PR04} is put to work by incrementally
constructing \(W\).
Our approach, being based on the dual check \(R \subseteq W^-\), adds
a new dimension of modularity/incrementality in which \(R\) is also
modified to safely exclude those transitions for which the current
proof is sufficient.
The advantage of the dual check was shown experimentally in
Sec.~\ref{sec:experiments}.
However, let us note that in our implementation we use potential
ranking functions and trace partitioning, which are not used
in ARMC~\cite{CookPR05}.
Moreover, 
it smoothly applies to conditional termination.

Kroening et al.~\cite{KroeningSTW10} introduced the notion of
compositional transition invariants, and used it to develop techniques
that avoid the performance bottleneck of previous
approaches~\cite{CookPR06}.
Recently, Chen et al.~\cite{ChenFM12} proposed a technique for proving
termination of single-path linear-constraint loops. 
%
Contrary to their techniques, we handle general transition relations
and our approach applies also to conditional termination.
As for conditional termination, the work of Cook et al.~\cite{CookGLRS08} is the closest to ours. However, we differ in the following points:
\begin{inparaenum}[\upshape(\itshape a\upshape)]
\item we do not use universal quantifier elimination, whose complexity
  is usually very high, depending on the underlying theory used to
  specify \(R\). 
  Instead, we adapt a fixpoint centric view that allows using abstract
  interpretation, and thus to control precision and performance;
  %
%
\item we do not need special treatment for loop with phase transitions
  (as the one of Sec.~\ref{sec:motivating}), they are handled
  transparently in our framework.
Bozga et al.~\cite{BozgaIK12} studied the problem of deciding
conditional termination. Their main interest is to identify family of
systems for which \(\gfp\ \lambda \mathcal{X}\ldotp
\pre[R](\mathcal{X})\), the set of non-terminating states, is
computable.

%
%
\end{inparaenum}

It is worth ``terminating'' by mentioning that several formulations, of the termination
problem, similar to the check \(R^+\subseteq W\) have appeared
before~\cite{CodishT99,LeeJB01,DershowitzLSS01}. They have also led to
practical tools for corresponding programming paradigms. The relation
between these approaches was recently studied~\cite{HeizmannJP10}.
Works based on these formulations, in particular those that construct
global ranking functions for \(R\)~\cite{Ben-Amram09}, might serve as
a starting point to understand some (completeness) properties of our
approach.
%
%
This is left for future work.

\bibliographystyle{splncs03}
\bibliography{ref}

%

\end{document}